\documentclass[prd,twocolumn,nofootinbib]{revtex4}
\usepackage{graphicx, epsfig}
\usepackage{color}
\usepackage{mathrsfs}
\usepackage {amssymb}
\usepackage {amsmath}

\newcommand{\nc}{\newcommand}
\newcommand{\nn}{\nonumber}
\nc{\ba}{\begin{eqnarray}}
\nc{\ea}{\end{eqnarray}}
\newcommand{\bea}{\begin{eqnarray}}
\newcommand{\eea}{\end{eqnarray}}
\newcommand\be{\begin{equation}}
\newcommand\beq{\begin{equation}}
\newcommand\ee{\end{equation}}
\newcommand\eeq{\end{equation}}

\nc{\s}{\sigma}
\newcommand\bk{\boldsymbol{k}}

\def\D{{\cal D}}
\def\R{\zeta}

\newcommand{\refeq}[1]{(\ref{#1})}

\def\e{{\tilde M}^2}
\def\M{M_P}
\def\c{c_{\cal D}}
\def\eps{\epsilon}
\def\invc{\left( \frac{1}{\c^2}-1\right)}
\def\matter{{\rm brane}}

\def\tt{{\tilde t}}

\def\z{\zeta}

\def\rc{\rho_c}

\def\teps{{\tilde \epsilon}}
\def\tH{{\tilde H}}
\def\tg{{\tilde g}}

\def\sd{{\dot \sigma}}
\def\X{\kappa}
\def\q{\alpha}

\def\w{w}

\def\ca{c_\sigma}
\def\ce{c_s}
\def\cae{c_{\s s}^2}

\def\l{\lambda}

\def\T{T_{\sigma s}}

\def\Sa{S^{(\s \s \s)}}
\def\Ss{S^{(\s ss)}}

\def\dQsi{\dot Q_{\sigma}}

\def\Qsi{Q_{\sigma}}
\def\pQsi{\partial Q_{\sigma}}
\def\LapQsi{\partial^2 Q_{\sigma}}

\def\dQs{\dot Q_{s}}

\def\pQs{\partial Q_{s}}

\def\kineticperp{ \dQs^2  }
\def\gradperp{ (\pQs)^2}

\input{epsf.sty}

\begin{document}

\title{Orthogonal non-Gaussianities from Dirac-Born-Infeld Galileon inflation}

\author{S\'ebastien Renaux-Petel}

\affiliation{Centre for Theoretical Cosmology, Department of Applied Mathematics and Theoretical Physics, University of Cambridge, Cambridge CB3 0WA, UK}


\begin{abstract}
We study the cosmology of the multifield relativistic Galileon model in which an induced gravity term is added to the Dirac-Born-Infeld action. 
We highlight the physical insight that is gained by employing a bimetric perspective in which the induced gravity and Einstein-Hilbert action are treated on equal footing.
We derive the conditions under which a phase of quasi exponential inflation can be sustained and demonstrate the existence of a critical background energy density above which cosmological fluctuations become ghosts. At the non-linear level, this scenario provides the first concrete early Universe model in which the shape of the bispectrum can be predominantly of orthogonal type. More generally, we show that the shape and sign of the primordial non-Gaussianities act as powerful discriminants of the precise strength of the induced gravity. 
\end{abstract}

\maketitle

The possibility that our Universe has more than four spacetime dimensions revolutionized theoretical cosmology in the past decade by bringing a geometrical perspective on two major issues. In the very early Universe, it entertains the idea that the inflaton(s) may be identified with the coordinate(s) of a brane in extra dimensions \cite{Dvali:1998pa}, resulting in completely new possibilities like the Dirac-Born-Infeld (DBI) scenario of inflation \cite{st03,Chen:2005ad}. In the late Universe, it suggested a higher-dimensional modification of gravity, like the Dvali-Gabadadze-Porrati model \cite{Dvali:2000hr} or more recently the Galileon \cite{Nicolis:2008in}, as a solution to the dark energy puzzle.

Following the geometrical picture behind these scenarios, namely the one of a D3-brane embedded in a five-dimensional spacetime, de Rham and Tolley demonstrated how DBI and the Galileon can be naturally generalized and unified \cite{deRham:2010eu}. This was further studied in the case of an arbitrary number of extra dimensions \cite{Hinterbichler:2010xn} based on \cite{Charmousis:2005ey}, with the result that the DBI Galileon model in an even number of extra dimensions strictly greater than two -- as relevant in string theory -- is geometrically surprisingly simple: in terms of the induced metric on the brane, its action is the sum of a cosmological constant (giving rise to DBI) and the corresponding Einstein-Hilbert term, to which the usual four-dimensional gravity is added. 

In this communication, we show how physically intuitive and computationally efficient it is to study the DBI Galileon scenario by treating the induced gravity and Einstein-Hilbert action on equal footing, which we call the \textit{bimetric} perspective. We derive the conditions under which this model can sustain a phase of quasi de-Sitter expansion, suitable for inflation or dark energy. We analyze the linear perturbations about a homogeneous cosmological solution and demonstrate the existence of a critical background energy density above which cosmological fluctuations become ghosts. Finally, we stress that this scenario 
provides the first concrete early Universe model in which the shape of the bispectrum can be predominantly of orthogonal type.

Our starting point is
\be
S= \int {\rm d}^4 x \left[ \frac{\M^2}{2}  \sqrt{-g} R[g]  +\frac{\e}{2} \sqrt{-\tg} R[\tg] + \sqrt{-g}  {\cal L}_\matter    \right]  
 \nn
\ee
where $g_{\mu \nu}$ is the cosmological metric and 
\be
\tg_{\mu \nu}=g_{\mu \nu}+fG_{IJ}\partial_{\mu} \phi^I \partial_{\nu} \phi^J
\label{induced}
\ee
is proportional to the induced metric on the brane (we simply refer to it as the induced metric in the following), $G_{IJ}(\phi^K)$ being a metric in the space of the scalar fields $\phi^I$ ($I=1,2 \ldots, N$), which correspond to the brane coordinates in the extra dimensions. The brane (DBI) action is given by ${\cal L}_\matter = -1/f ( \sqrt{\D}-1 )-V(\phi^I)$ where the explicit expression of ${\cal D} \equiv {\rm det} (g^{-1}\tg)$ in terms of the scalar fields was given in \cite{Langlois:2008wt}. Here, we consider a constant warp factor $f$ for simplicity of presentation, so that the mass scale $\e$ can be considered as constant, leaving the analysis of the general case to a companion paper \cite{RenauxPetel:2011uk}.

One can express $ \sqrt{-\tg}R[\tg]$ in terms of the fields and the geometrical quantities associated with the cosmological metric, leading to a multifield relativistic extension of the quartic Galileon Lagrangian in curved spacetime. The resulting expression is very intricate \cite{RenauxPetel:2011uk} and rather obscures the physics which is at play by breaking the symmetry between the cosmological and the induced metric in the ``gravitational'' part of the action. Here, we rather keep this symmetry manifest and hence write the (gravitational) equations of motion in the compact form
\bea
\label{metric-eoms}
 \M^2  G^{\mu \nu}[g]+\e G^{\mu \nu}[\tg] \sqrt{\cal D}=T^{\mu \nu}_\matter \,. 
\eea

Now specializing to a flat Friedmann-Lema\^itre-Robertson-Walker (FLRW) background spacetime in which the scalar fields are only time-dependent, let us highlight the main properties of the homogeneous evolution. First, like in standard brane inflation, it is useful to
introduce the background value of $\D$: $\c^2 \equiv 1- f  \dot \sigma^2$, where $\sd \equiv \sqrt{G_{IJ} \dot \phi^I   \dot \phi^J}$ 
plays the role of an effective collective velocity of the fields. The value of $\c^2$ distinguishes the slow-roll regime where $\c^2 \simeq 1$ from the so-called relativistic, or DBI, regime, where the brane almost saturates its speed limit and $\c^2  \ll 1$. Let us then note that in the background, the induced metric \refeq{induced} takes the form of a flat FLRW metric whose scale factor is the cosmological one, $a$, but whose cosmic time $\tt$ is such that $d \tt=\c dt$. Introducing the corresponding Hubble and ``deceleration'' parameters, given respectively by $\tH \equiv  \frac{1}{a} \frac{d a}{d \tt} = \frac{H}{\c}$ and $\teps  \equiv  -\frac{1}{\tH^2} \frac{d \tH}{d \tt} =   \eps + s$, where $\eps \equiv - \dot H/H^2$ and $s \equiv \dot{c}_{\D}/(H \c)$, it is then straightforward to derive the modified Friedmann equations
\bea
\label{Friedmann1}
 3 H^2 \left(\M^2 +\frac{\e}{\c^3} \right)=\rho_\matter = V+ \frac{1}{f}\left( \frac{1}{\c}-1 \right)
\eea
\bea
\M^2 H^2 \eps +\frac{ \e H^2}{\c} \left( \eps +s+\frac32 \invc \right)= \frac{\sd^2}{2\c}\,.
\label{Hdot}
\eea
From the third term in parenthesis in \refeq{Hdot}, one deduces that for generic values of $\e$, $\eps$ acquires a large negative contribution when $\c^2 \ll1$. Quantitatively, one can show \cite{RenauxPetel:2011uk} that barring cancellations, achieving a phase of quasi de-Sitter expansion $\eps \ll 1$ in the relativistic regime requires, in addition to the usual condition to enter into the DBI regime $\c f V \gg 1$, that $\e  \ll \M^2 \c^3$. Despite this restrictive condition, the induced gravity can still have a non negligible effect on the background evolution, as can be seen for example through the approximate equation deduced from \refeq{Hdot} (neglecting $s$): $\eps \simeq 3/2 (\c f  V)^{-1}  (1-3\q)$, where the dimensionless quantity 
$\q \equiv f H^2 \e/\c^2$
can be of order one. Note however that this contribution to $\eps$ is always positive in the regime where the theory is ghost-free, as demonstrated by the analysis of cosmological perturbations, to which we now turn.

In order to study the dynamics of linear perturbations about a homogeneous cosmological solution, we use the ADM formalism in which the metric is written in the form
\beq
g_{\mu \nu} dx^{\mu} dx^{\nu}=-N^2 dt^2 +h_{ij} (dx^i+N^i dt)(dx^j+N^j dt)\,
\eeq
where $N$ is the lapse function and $N^i$ the shift vector.
Following the bimetric perspective, it proves to be useful to introduce their induced gravity counterparts and to write the ``gravitational'' part of the action as two copies of the Einstein-Hilbert action in the ADM form. It is then relatively straightforward to solve the constraint equations for $N$ and $N_i$ perturbatively, insert their solutions in the action and deduce the quadratic action in terms of the propagating degrees of freedom: two tensor modes and $N$ scalar degrees of freedom. 

Before discussing the multifield situation, it is instructive to consider the case of a single inflaton field $\phi$ ($N=1, G_{11}=1$). One can then choose the uniform inflaton gauge in which the scalar perturbation $\z$ appears in the spatial metric $h_{ij}$ in the form $h_{ij}=a(t)^2 e^{2 \z} \delta_{ij}$ while the inflaton is homogeneous $\phi=\phi(t)$. The resulting second-order (scalar) action can then be cast in the simple form
\bea
S_{(2)}= \int {\rm d}t\,    {\rm d}^3 x \,a^3  \left(  A(t) {\dot \zeta}^2 -B(t) \frac{(\partial \zeta)^2}{a^2}   \right)
\label{2d-order-action-2}
\eea
where
\bea
A(t)&=& \frac{\M^2}{\c^2} \left( \eps \, \X^2+3\c^2(1-\X^2) \right) \nn \\
&+& \frac{\e}{\c^3} \left( \teps \, \X^2+3\c^2\left(1-\frac{\X^2}{\c^4}\right) \right) \,,\label{A}  \\
B(t) &=& \M^2 \left( \eps \, \X (3 \, \X-2)+\X-1  \right)  \nn \\
&+&\frac{\e}{\c} \left( \teps \, \X \left(3 \,\frac{ \X}{\c^2}-2\right)+\X-\c^2 \right)\,,
\eea
with $\X \equiv  \left( \M^2+\frac{\e}{\c}  \right) /  \left( \M^2+\frac{\e}{\c^3} \right)$. This formulation reveals how the fluctuation $\zeta$ reacts to the two background geometries: without the induced gravity, $\e=0$, $\X=1$ and one recovers the standard $k$-inflationary result \cite{Garriga:1999vw} with a speed of sound $B(t)/A(t)=\c^2$. At the other extreme limit, if one formally considers $\M^2=0$, then $\X=\c^2$ and one finds a similar expression
\bea
S_{(2),\M^2=0}= \int {\rm d} \tt\, {\rm d}^3 x \,a^3 \, \teps \, \e \c^2  \left(   \left(\frac{d \zeta}{d \tt}\right)^2 -\frac{1}{\c^2} \frac{(\partial \zeta)^2}{a^2}   \right) \nn
\eea
in terms of the quantities associated with the induced metric with the replacement 
$\c^2 \to \c^{-2}$. This can be easily understood: one can then treat the metric $\tg$ as the cosmological metric, in which case the action for gravity becomes canonical while the brane action can be expressed as $S_\matter =  \int d^4 x \sqrt{-\tg} P(X_{\tg},\phi)$ where $X_{\tg} \equiv -\frac{1}{2}\tg^{\mu \nu}\partial_{\mu} \phi \partial_{\nu} \phi $ and
\bea
P= \left( \frac{1}{f}-V \right) \sqrt{1+2 f X_{\tg}}  -\frac{1}{f}\,.
\label{Pmodified}
\eea
Using the general expression derived in \cite{Garriga:1999vw}, it is easy to verify that the speed of sound is given by $1/\c^2$  for the Lagrangian \refeq{Pmodified}, with the consequence that the speed of sound with respect to the cosmological cosmic time is unity. In the general situation, one can not define an Einstein frame in which the gravitational action becomes canonical and one must resort to the full calculation, leading to the result \refeq{2d-order-action-2} that nicely interpolates between the two extreme cases aforementioned. However, the form \refeq{A} of the kinetic term is not appropriate for discussing the possible presence of a ghost. Because the conditions for avoiding the presence of ghosts are, \textit{a priori}, more restrictive in the multifield case, we now turn to this situation, restricting our attention to two fields ($I=1,2$) for simplicity of presentation. 

It is then convenient to express the two scalar degrees of freedom in terms of the scalar field perturbations in the flat gauge, that we denote $Q^I$, and to decompose them into $Q^I=Q_\s e^I_\s+Q_s e^I_s$\,, where $e^I_{\s} \equiv \frac{ \dot \phi^I}{\sd}$ is the unit vector pointing along the background trajectory in field space and $e_s^I$ is the unit vector orthogonal to $e^I_\s$. The so-called adiabatic perturbation $Q_{\s}$ inherits all the properties of the singe field case while $Q_s$, called the (instantaneous) entropy perturbation, embodies the genuinely multifield effects. Remarkably, after going to conformal time $\tau = \int {{\rm d}t}/{a(t)}$, and, upon using the canonically normalized fields $v_{\s} \equiv   z\frac{H}{\sd} \, Q_{\s}$ and $v_{s} \equiv  \w \, Q_s$ with 
\bea
&&z= \frac{a}{\c^{3/2}} \frac{\sd}{H} \left((1-9 \q)\X^2+6 \q \X \c^2  \right)^{1/2} \label{z} \,,\\
&&\w=\frac{a}{\sqrt{\c}}\,\left(1-3\q\right)^{1/2}  \label{alpha}\,,
\eea
one can express the exact second-order action, and hence the equations of motion, in a very compact form similar to the one in standard multifield DBI inflation \cite{Langlois:2008wt}:
\begin{eqnarray}
\hspace*{-1.5em}
 &&v_{\s}''-\xi v_{s}'+\left(\ca^2 k^2-\frac{z''}{z}\right) v_{\s} -\frac{(z \xi)'}{z}v_{s}+\cae k^2 v_s=0\,,  \nn \qquad
\label{eq_v_sigma}
\\
\hspace*{-1.5em} 
 &&v_{s}''+\xi  v_{\s}'+\left(\ce^2 k^2- \frac{\w''}{\w}+a^2\mu_s^2\right) v_{s} - \frac{z'}{z} \xi v_{\s}+\cae k^2 v_{\sigma}=0 \nn \,,
\label{eq_v_s}
\end{eqnarray}
where 
\begin{eqnarray}
\label{11}
&&\ca^2 =  \left(  (1-9 \q)\X+6 \q \c^2  \right)^{-1}  \nn \\ 
&&\times  \left[ \frac{\c^2}{\X} \left( (1-5 \q)\X+2 \q \c^2 \right)  +6 \q s \c^2 \frac{\X-\c^2}{1-\c^2}    \right]  \,,
\\
&&\ce^2= \frac{\c^2}{\X} \left[1+ \frac{2 s \q}{1-3 \q}  \frac{\X-\c^2}{1-\c^2}    \right]  
\label{z,a}
\end{eqnarray}
and the explicit expressions of $\mu_s^2, \xi$ and $\cae$ will be presented elsewhere \cite{RenauxPetel:2011uk}. By taking the square roots in \refeq{z}-\refeq{alpha}, we implicitly consider the regime in which the fluctuations are not ghosts, which reads $\q < (9-6 \c^2/\X)^{-1}$. Interestingly, note that this condition on $\q$ can be reformulated, through the first Friedmann equation \refeq{Friedmann1}, as an upper bound on the energy density $\rho_\matter$, \textit{i.e.} there exists a critical background energy density $\rc =  \left(1+\frac{\c^3 \M^2}{\e}\right) /    \left( f \c \left(3-2\frac{\c^2}{\X}\right) \right)$
above which the cosmological fluctuations become ghosts. 
Incidentally, note that the naive procedure consisting in only perturbing the scalar fields (and not the metric) would give a completely wrong second-order action as soon as $\X$ differs significantly from 1 \cite{RenauxPetel:2011uk}.

In the remaining of this communication, we consider a quasi de-Sitter inflationary phase in a slow-varying regime in which the time evolution of every quantity is slow with respect to that of the scale factor -- we neglect $s$ in particular -- so that $\ca^2$ and $\ce^2$ are positive definite. A novel feature, due to the common presence of both the Einstein-Hilbert and induced gravity action, is the gradient coupling between adiabatic and entropy perturbations through the term $\cae$. We will assume that the effect of this coupling, as well as the one coming from $\xi$, can be neglected on sub-horizon scales, so that the adiabatic and entropy perturbations can be quantized independently. In that case, $\ca$ and $\ce$  are truly their respective speed of propagation, which are in general \textit{different}: the entropic speed of sound is always less than the adiabatic one, the two becoming equal when $\q \to 0$ for any $\c$ (this is standard multifield DBI inflation) and when $\c \to 1$ for any $\q$ (the non-relativistic limit of the DBI Galileon). Finally, note that one should impose the more stringent condition $\q \leq  (9-2\c^2/\X)^{-1}$ if we additionally require the sound speeds to be less than one. 

Under the assumptions above, it is easy to deduce the power spectrum of the adiabatic and entropy fluctuations \cite{RenauxPetel:2011uk}. As for the observable comoving curvature perturbation $\z=-(H/\sd) Q_\s$, its late-time power spectrum can be formally written as ${\cal P}_{\R}= {\cal P}_{\R_*}   \left( 1+\T^2 \right)$ where the transfer coefficient $\T$ is due to the feeding of the adiabatic perturbation by the entropic perturbation on super-horizon scales \cite{Wands:2002bn} and the power spectrum around adiabatic sound horizon crossing is given by
\be
{\cal P}_{\z_*}= \left( \frac{H}{\sd} \right)^2  \left( \frac{H}{2 \pi} \right)^2 \frac{  \left((1-9 \q)\X^2+6 \q \X \c^2  \right)^{1/2}     }{ \left((1-5 \q)\X+2 \q \c^2  \right)^{3/2}  } \,.
\label{power-spectrum-R-*}
\ee

We finally discuss the impact of the induced gravity on the primordial non-Gaussianities generated in the relativistic regime $\c^2 \ll 1$. It is then sufficient to perturb only the scalar fields and, following the same bimetric approach as for the calculation of the quadratic action, the dominant contribution to the third-order action is found to be \cite{RenauxPetel:2011uk}
\bea
S_{(3)}^{\rm eff}&=& \int {\rm d}t \,{\rm d}^3 x\,   \frac{a^3}{2 \c^5 \sd} \left[ A_{\dQsi^3}   \dQsi^3 
 - \c^2 A_{\dQsi (\pQsi)^2}  \dQsi \frac{(\pQsi)^2}{a^2}
\right.
\cr
&&
\left.
\hspace*{-4.0em}+ \c^2 \dQsi \kineticperp+\c^4 \dQsi \frac{(\pQs)^2}{a^2}-2 \c^4 (1-2\q) \frac{\pQsi \pQs}{a^2} \dQs  
\right.
\cr
&&
\left.
\hspace*{-4.0em} -\frac32 \q \c^4\left( \frac{1}{\l^2}-1 \right)  \kineticperp \frac{\LapQsi}{H a^2}
+ \frac{\q}{2} \c^6\left( \frac{1}{\l^2}-1 \right) \frac{\gradperp}{a^2}  \frac{\LapQsi}{H a^2}
\right] \nn
\label{S3}
\eea
where $A_{\dQsi^3} =1-3\q \left(5-4 \l^2+\l^4 \right)$, $A_{\dQsi  (\pQsi)^2} = 1- \q \left( 9-3 \l^2 \right)$ and $\l^2 \equiv \ce^2/\ca^2 \simeq \frac{1-9\q}{1-5\q} $
in the slow-varying relativistic regime. Here, we used the linear equation of motion for $\Qsi$ to trade the higher dimension operators $\dQsi^2 \partial^2 \Qsi$ and $(\partial \Qsi)^2 \partial^2 \Qsi$ for the usual $k$-inflationary ones $\dQsi^3$ and $\dQsi (\pQsi)^2$ \cite{Mizuno:2010ag,Creminelli:2010qf,RenauxPetel:2011sb}. The same is only partially possible for the mixed vertices including adiabatic and entropy perturbations, which explains the appearance of the last two higher dimension vertices. Following the standard procedure \cite{Maldacena:2002vr}, one can compute the associated primordial bispectrum -- the three-point correlation function of $\z$ at late time -- concentrating on the contributions induced by the quantum interactions around horizon crossing. This reads
\be
\langle \R(\bk_1) \R(\bk_2) \R(\bk_3) \rangle = (2 \pi)^7 \delta (\sum_{i=1}^3 \bk_i) {\cal P}_{\R}^2 \frac{S(k_1,k_2,k_3)}{(k_1 k_2 k_3)^2} \nn
\ee
where the shape function
\bea
S=-\frac{1}{2 (1+\T^2)^2} \frac{1}{\c^2 (1-9 \q)} \left( \Sa+ \T^2 \Ss \right) \nn
\eea
has been separated out between its purely adiabatic and entropy-induced components. The adiabatic shape
\bea
\Sa =3 A_{\dQsi^3} S_{\dQsi^3}+ \frac{\l^2}{2}A_{\dQsi  (\pQsi)^2}  S_{\dQsi (\pQsi)^2}
\eea
is a linear combination (function of $\q$) of $S_{\dQsi^3}=(k_1 k_2 k_3)/K^3$ and $S_{\dQsi (\pQsi)^2}=-k_3/(k_1 k_2 K^3) ( \bk_1 \cdot \bk_2) \left(2 k_1 k_2 -k_3 K+   2K^2 \right)+ 2\,{\rm perm.}$ where $K \equiv  k_1+k_2+k_3 $ and the `perm.' indicate two other terms with permutations of indices 1, 2 and 3. To assess the experimental ability to distinguish between different bispectrum momentum dependence, we use the scalar product between shapes introduced in \cite{Fergusson:2008ra}. In this sense, $S_{\dQsi^3}$ and $S_{\dQsi (\pQsi)^2}$ are very similar -- in particular, they peak on equilateral triangles -- which explains why they are often approximated by a common ansatz called equilateral \cite{Creminelli:2005hu}. They are different though \cite{Chen:2006nt} and, as pointed out in \cite{Senatore:2009gt}, one can highlight their differences by considering an appropriate linear combination of them with respect to which it is almost orthogonal. These \textit{orthogonal} non-Gaussianities are now routinely considered in cosmic microwave background data analysis, with the constraint $ -410 < f_{NL}^{orth} < 6\, (95 \% \, {\rm C.L})$ \cite{Komatsu:2010fb} on their amplitude $f_{NL}^{orth} \equiv \frac{10}{9}S_{orth}(k,k,k)$. However, a concrete early Universe model that generates such a non-gaussian signal was still lacking. 
\begin{figure}[t]
\includegraphics[width=0.23\textwidth]{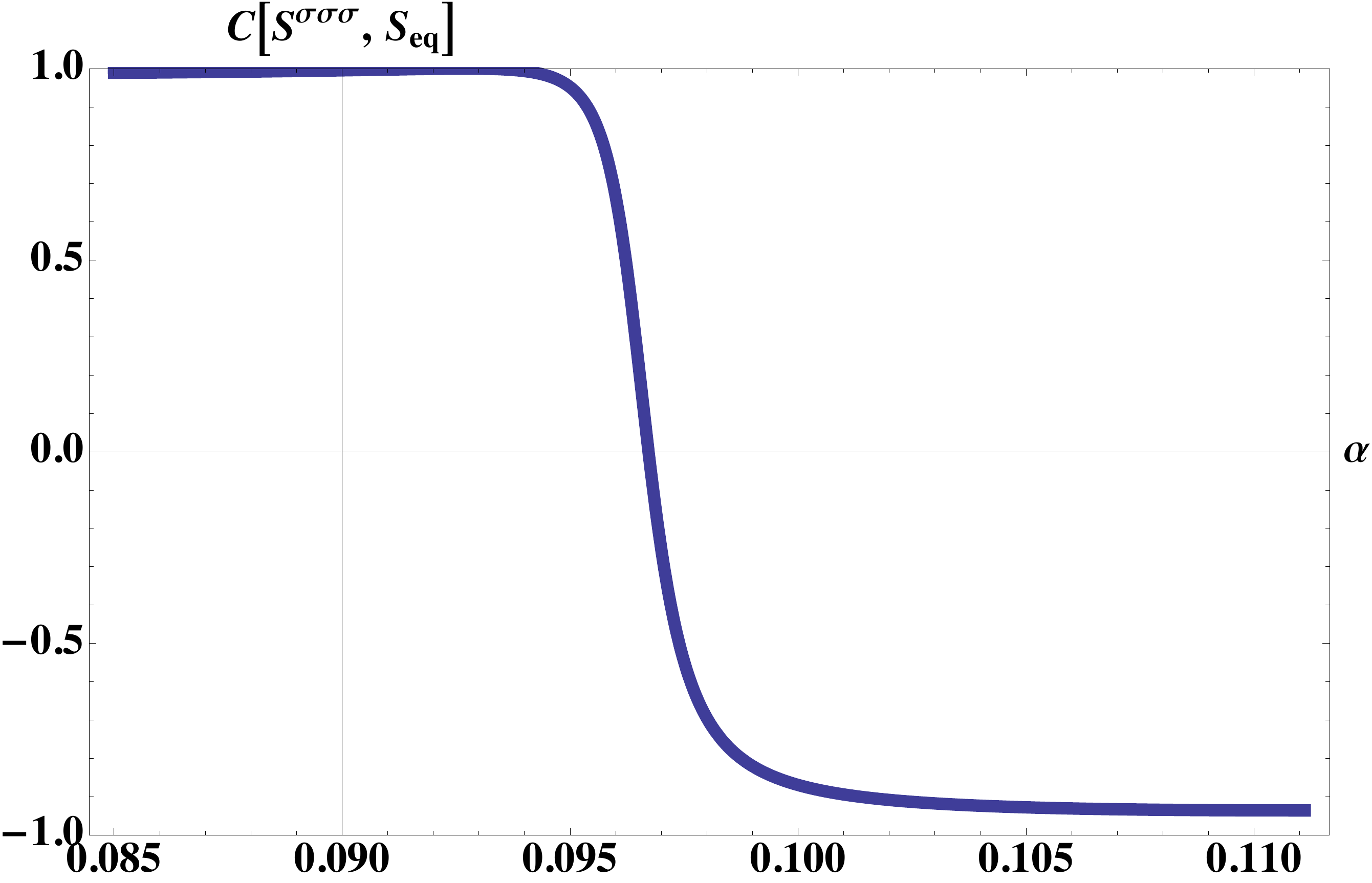}
\space
\space
\includegraphics[width=0.23\textwidth]{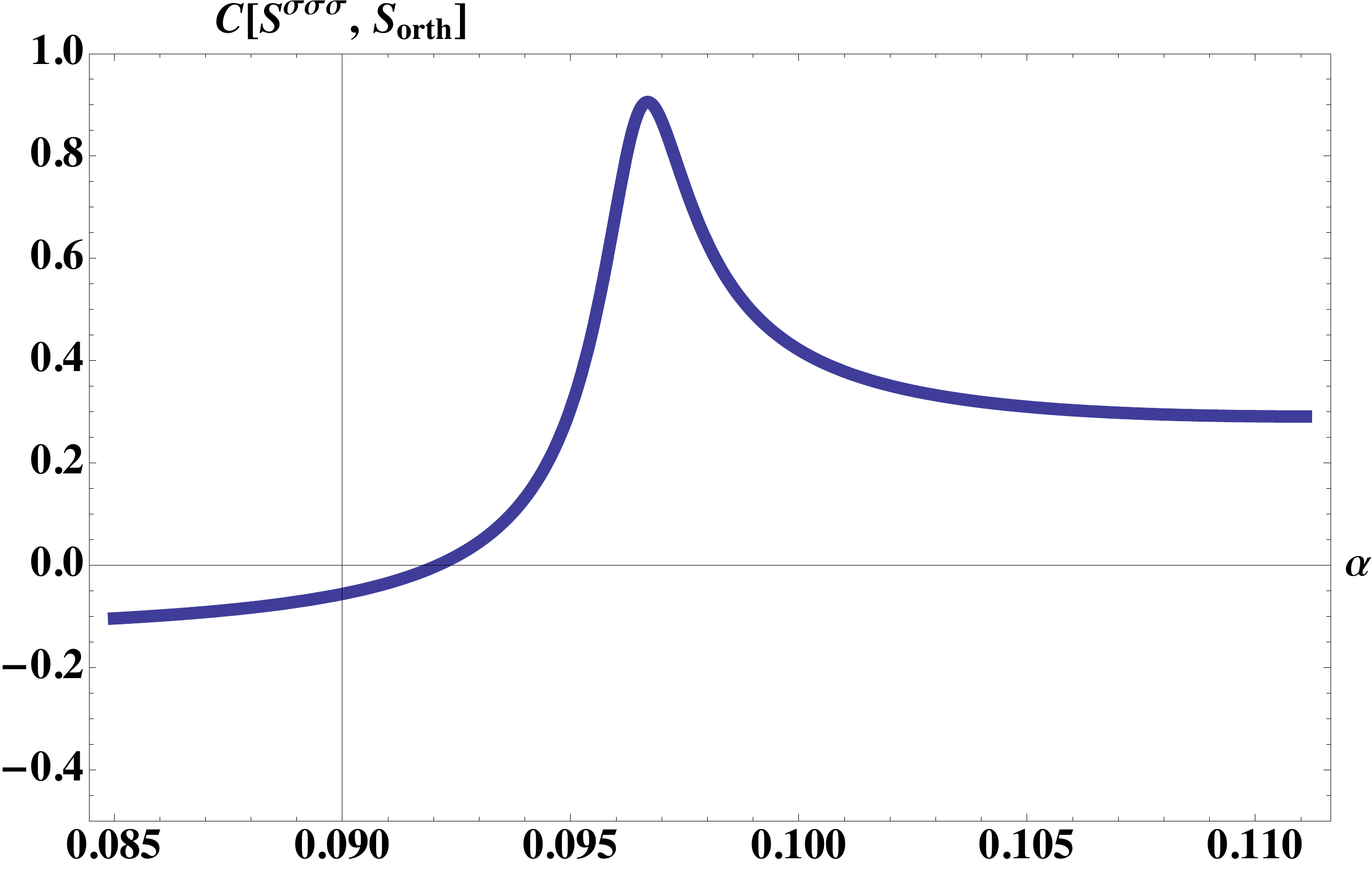}
\caption{\label{correlation} {\small Correlation of the adiabatic shape $S^{(\s \s \s)}$ with the equilateral ansatz (left) and the orthogonal ansatz (right) as a function of $\q$.}}
\end{figure}
Figure \ref{correlation} demonstrates that DBI Galileon inflation provides such a model: as $\q$ increases, the correlation between the adiabatic shape and the equilateral ansatz decreases from $1$ to $-1$ (left) and orthogonal non-Gaussianities are generated in the narrow transient region (right). The maximum correlation equals 91 \% and is reached at $\q \simeq 0.097$, for which we find $f_{NL}^{orth} \simeq -\frac{0.016}{\c^2}$. The entropy-induced shape is different but qualitatively similar to the adiabatic one: it interpolates between negative (when $\q=0$) and positive equilateral non-Gaussianities passing through orthogonal ones, although for smaller values of $\q$ centered around $0.08$ \cite{RenauxPetel:2011uk}.

To summarize, we have studied the cosmology of the relativistic Galileon model in which an induced gravity term is added to the DBI action governing the motion of a D3-brane in a higher-dimensional spacetime. By treating it as a genuine modification of gravity, we have been able to  analyze it in a physically transparent and computationally very efficient way. The induced gravity tends to violate the null energy condition and to render cosmological fluctuations ghosts. There nonetheless exists an interesting parameter space in which a stable phase of quasi-exponential expansion can be achieved while the induced gravity leaves non-trivial imprints on the cosmological fluctuations. In particular, this scenario 
provides the first concrete early Universe model in which the shape of the bispectrum can be predominantly of orthogonal type. More generally, the fact that qualitative aspects of the latter, such as its shape and sign, depend non-trivially on the precise value of the induced gravity strength, exemplifies the usefulness of higher-order statistics for pinning down the mechanism that seeded the large-scale structure of the Universe.

{\it Acknowledgments:} I thank S. Mizuno and K. Koyama for collaboration on this subject and X. Chen, D. Langlois and R. Ribeiro for useful comments on a preliminary version of this manuscript. I am supported by the STFC grant ST/F002998/1 and the Centre for Theoretical Cosmology.


\end{document}